# On-chip spectrometers using stratified waveguides filters

ANG LI,[1,*] AND YESHAIAHU FAINMAN[1]

[1]Department of Electrical and Computer Engineering, University of California at San Diego, La Jolla, California 92093, USA
*angli@ucsd.edu

Abstract

We present an ultra-compact single-shot spectrometer on silicon platform with broad operation bandwidth and high resolution. It consists of 32 stratified waveguide filters (SWFs) with diverse transmission spectra for sampling the unknown spectrum of the input signal and a specially designed ultra-compact structure for splitting the incident signal into 32 filters with low imbalance. Each SWF has a footprint less than 1um x 30um, while the 1x32 splitter and 32 filters in total occupy an area of about 35um x 260um, which to the best of our knowledge, is the smallest footprint spectrometer realized on silicon photonic platform. Experimental characteristics of the fabricated spectrometer demonstrate a broad operating bandwidth of 180nm centered at 1550nm and narrowband peaks with 0.45nm Full-Width-Half-Maximum (FWHM) can be clearly resolved. This concept can also be implemented using other material platforms for operation in optical spectral bands of interest for various applications.

## 1. Introduction

Optical spectrometers are indispensable elements in material science, chemical sensing, astronomical science, and in-situ medical applications. The state-of-the-art high-performance spectrometers are currently realized by bulky, high cost systems, thereby, limiting their deployment in various system applications. For example, miniaturized spectrometers could be integrated into intelligent portable devices, such as smartphones and unmanned aerial vehicles, in support of overarching internet of things applications. Among various approaches to miniaturized spectrometers, silicon photonics is most promising due to its compatibility with low-cost CMOS manufacturing technology(1). Additionally, silicon photonics approach provides ultra-high index contrast that allows extremely compact device footprint in support of high integration density (1) and broad transparent spectral bandwidth window between 1.1 um to 2.2 um. This approach can be further extended by employing other CMOS compatible materials such as SiN (2-4) with low propagation loss of about 1 dB/cm, enabling efficient implementation of long delay lines which are essential for achieving high spectral resolution in various types of integrated spectrometers(5, 6). Moreover, with hybrid integration approach, it is possible to integrate high performance light source (7-9), high-speed photodetectors(10, 11), electronic driver and electronic signal processing circuits in a single chip-based package, leading to integrated miniaturized spectrometer systems for various practical applications(12, 13).

Numerous approaches have been carried out to realize high performance spectrometers integrated on silicon platform. The most straightforward approach is to disperse the spectral content of the incident signal into an array of photodetectors and record the corresponding spectral signals (14-20). The dispersion can be achieved by an array of narrowband filters or dispersive elements such as gratings as shown in Fig.1 (a-b). Although these types of spectrometers could theoretically operate with broad spectral bandwidth signals and provide high spectral resolution, it will be achieved at the expense of introducing very large number of building blocks (such as grating channels or filters and high-performance photodetectors), which in turn will result in considerably large footprint, high insertion loss, hardware cost and operation complexity. For instance, echelle grating based spectrometer with a footprint as large as 3x3 mm$^2$ can produce a resolution of $\delta\lambda$=0.5 nm, but it requires a large number of channels $N$ (e.g., over 120) to achieve moderate wavelength bandwidth of $\Delta\lambda$=60 nm since the frequency bandwidth, $\Delta\omega = N\delta\omega$ which reduces to $\Delta\lambda = N\delta\lambda$ (18). Similar issues apply to ring resonators array-based spectrometer, which employs as many as 84 ring resonators to produce a resolution of 0.6 nm supporting the total bandwidth of 50 nm(19). Both performance indicators are non-trivial to improve since the bandwidth is fundamentally limited by the free spectral range (FSR) of silicon ring resonators whereas the resolution is directly proportional to the quality factor, Q (3 dB bandwidth) of the individual ring resonator. Moreover, for ring resonator with high Q factor, the backscattering induced resonance-splitting will emerge and significantly deteriorate the spectrometer performance(21). However, most critical concern of this class of spectrometers is their poor signal to noise ratio (SNR). Since the incident signal is divided into multiple narrow band signals to achieve the required spectral resolution, the power in each detection channel, linearly proportional to the ratio of bandwidth over the resolution, will be affected by the finite dynamic range and SNR of each corresponding photodetectors leading to poor SNR in the reconstructed broadband spectrum of the input signal, especially for the spectral components with very low magnitude. Another drawback lies in the poor tolerance to inevitable fabrication imperfections, which would introduce phase errors to multi-channel gratings (such as Arrayed Waveguide Grating(14)) and unpredictable resonance shifts to ring resonators, both of which are detrimental to the spectrometers' performance.

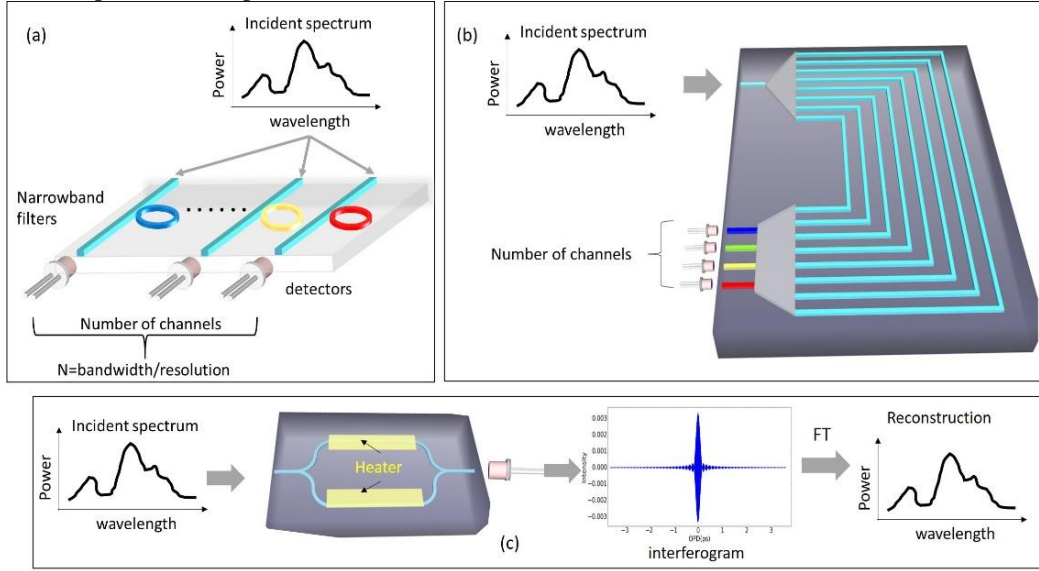

Fig. 1 Conceptual illustrations of different types of spectrometers. Schematic diagram describing the conventional split-and-detect spectrometers consisting of (a) narrowband filters array or (b) dispersive grating for fast spectrum reconstruction. The number of required channels is linearly proportional to the ratio of reconstruction bandwidth and resolution. The Fourier transform spectrometer (FTS) implemented on silicon photonics platform shown conceptually in (c) is used to generate and detect an interferogram (i.e., autocorrelation of the input signal displayed by left part of (c)) by changing optical path delay (OPD) between two arms of a balanced Mach-Zenhder-Interferometer (MZI) using integrated heaters. The spectrum of the input signal on the right part of (c) can be retrieved by performing Fourier transformation (FT) of the interferogram.

In contrast, due to Fellgett's advantage in terms of high SNR(22), Fourier transform spectrometers (FTS) have been extensively investigated (23, 24). The FTS principle of operation is based on detection of the auto-correlation function, or in other words, temporal interferogram of the incident signal by continuously tuning the optical path delay (OPD) between two pathways of the incident signal in an interferometer, and measuring the interference intensity at each OPD as illustrated in Fig.1(c). This type of spectrometer has been realized on silicon photonic platform using a Mach-Zenhder-Interferometer (MZI) structure with integrated heaters used to change the waveguide effective index and, consequently, the OPD between the two arms of the MZI. This approach demonstrated operation with broadband input spectral signals (i.e., ~ 60 nm) and high resolution (i.e., sub nm)(24), however, it was achieved at the expense of high power consumption of the heaters (> 5 W), large driving voltage (~ 160 V) and long measurement time (> 1.5 hours) due to slow thermo-optic effect. Moreover, the footprint of this realization is still considerably large (i.e., > 1 mm$^2$).

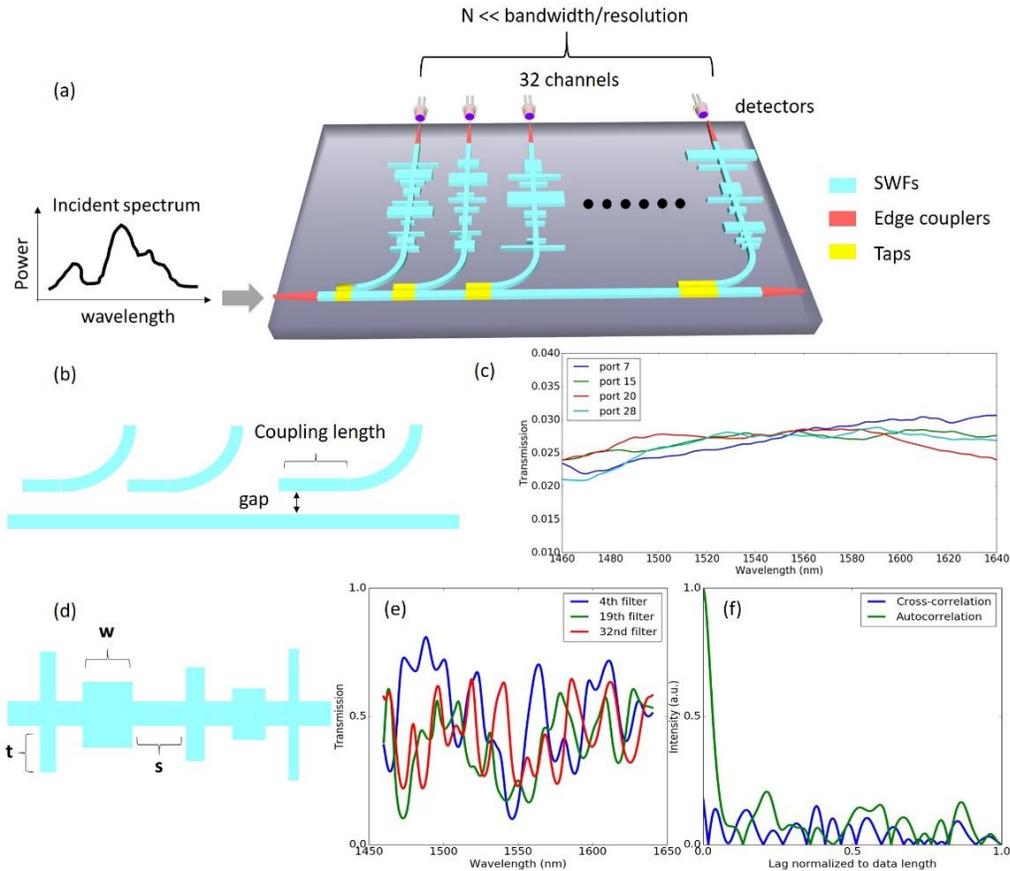

Fig.2 Illustration of single-shot spectrometer enabled by broadband stratified waveguides filters (SWFs): (a) depicts the simplified schematic of our implementation of this system on silicon. It contains two core parts: a 1x32 splitter based on cascaded taps and 32 broadband stratified waveguides filters (SWFs) with ultra-compact footprint and diverse spectral features. The number of channels needed is much smaller than in the split-and-detect concepts in Fig. 1 (a and b); (b) illustrates the structure for the cascaded taps with (c) showing the simulated transmission spectra at some output ports of the taps; (d) description of the SWF design with the defined design degrees of freedom s, w, and t to control the transmission properties of the stratified waveguide; (e) presents the simulated transmission spectra of some of the designed SWFs and (f) shows the auto-correlation of the spectrum of a single SWF as well as the cross-correlation of two distinct SWFs.

Alternatively, a set of broadband filters with diverse spectral responses can be employed to sample the incident signal as schematically illustrated in Fig. 2(a)(25, 26). With a much smaller number of channels compared with the split-and-detect spectrometers, the incident spectrum can be reconstructed instantaneously using computationally efficient signal processing algorithm. Intuitively, this approach is expected to have much smaller than FTS footprint and simultaneously provide a higher SNR in comparison to the split-and-detect spectrometers approaches. In addition, the set of broadband filters approach not only reduces the footprint compared to FTS, but also does not require electrical driving/heating and long sampling time. The key challenge for the set of broadband filters approach is to develop a series of broadband filters with very different, ideally orthogonal, spectral response. Recent demonstrations of this type of single-shot spectrometers utilize either 195 colloidal quantum dots (CQD) absorption filters (25) or 36 photonics crystals (PhC) cavities on Silicon-On-Sapphire substrate (26). Both approaches have good performance (a few hundred nm of bandwidth with 1-2 nm resolution), however, they do have individual challenges that are non-trivial to overcome. For the CQD based approach (25), the biggest challenge lies in the difficulty to integrate CQD into a CMOS compatible material platform supporting massive fabrication capability. Moreover, due to the similar filter response, a large number of filters (195) is needed to reconstruct a 300nm bandwidth with 2nm resolution. Additionally, the incident signal has to be split into 195 channels, which is even worse in terms of SNR compared to the split-detect approaches to achieve single-shot spectrum reconstruction. For the PhC cavities approach (26), the incident light is coupled from free space modes into slab cavity modes experiencing an appreciable mode mismatch between a 210x210 um² aperture with corresponding free-space modes and highly confined photonics crystal modes, resulting in a considerable coupling loss. Moreover, the performance of the spectrometer is very sensitive to the variation of the free space modes incidence angle, which, in practice, is difficult to accurately control. Additionally, the reported configuration is integrated with a commercially available CMOS sensor, resulting in a large device volume and slow response.

In this manuscript, we propose and demonstrate experimentally a chip-scale single-shot spectrometer using stratified waveguides filters (SWF) on silicon platform. Due to the rich design degrees of freedoms, the SWFs can be constructed to produce spectral responses with very low cross-correlation, leading to high sampling accuracy of the incident signal. Consequently, only 32 filters are adequate for detection and reconstruction of broad spectral bandwidth signals with good quality. In our experimental realization, each SWF has a footprint less than 1x30 um², much smaller compared with the work in (26). Additionally, we also construct an ultra-compact uniform splitter to introduce the incident signal into the 32 SWFs. The core part of the 1x32 splitter and the 32 filters occupy a total area of about 35x260 um², which to the best of our knowledge, is the smallest footprint spectrometer implemented in silicon photonic platform. The fabricated SWF-based spectrometer was interfaced with an optical fiber with low coupling loss and characterized using a broadband, a narrow band and a combination of broad-and narrow-band input signal. The measured experimental results demonstrate operation with both broadband (>100 nm) and narrowband (0.45 nm FWHM) spectrum signals.

## 2. Principle and system design

The detected power at a photodetector of an unknown signal with spectrum power $P(\lambda)$ passing through a broadband filter with transmittance $F(\lambda)$, can be mathematically written as:

$$D = \int P(\lambda)F(\lambda)d\lambda \qquad (1)$$

where $D$ represents the detected power of the transmitted input signal. The transmission spectrum of the filter $F(\lambda)$ can be accurately measured during calibration process. Ideally, we are dealing with continuous variables, however, during the reconstruction processes we will reconstruct the digitized values of the input signal power spectrum. Therefore, for simplicity we represent digitized values of both $P(\lambda)$ and $F(\lambda)$ by 1-D signals, i.e., vectors with values $P(\lambda_m)$ and $F(\lambda_m)$, $with\ m = 1,2\ldots M$. The length of these vectors described by number M determines the spectral resolution of the reconstructed input signal. With this transformation, equation (1) can be rewritten for the n-th filter as:

$$D_n = \sum_{m=1}^{M} P(\lambda_m)F_n(\lambda_m) \qquad (2)$$

Clearly, with N distinct filters, we will generate N corresponding values of $D_n$ with n=1, 2, …N. This formulation provides N linear algebraic equations that can be solved to determine M unknown values of the input signal, $P(\lambda_m)$. In a linear algebra formulation, the responses of N filters yield,

$$D_{Nx1} = S_{NxM}P_{Mx1}, \qquad (3)$$

where a sampling matrix $S_{NxM}$ of size [NxM] connects the detected intensities vector $D_{Nx1}$ of length N with input signal vector $I_{Mx1}$ of length M. With a proper design, the number of filters required for reconstruction of the input signal can be much smaller than the ratio of the targeted bandwidth over the desired spectral resolution, i.e. N<<M. Therefore, it outperforms previous split-and-detect spectrometers in multiple aspects including dynamic range, SNR, footprint, hardware cost and system operation complexity. In comparison with FTS techniques, this approach has the advantages of no power consumption and ability to instantaneously reconstruct the spectrum on the expense of reduced SNR. Note that, the rank of matrix S needs to be as large as possible, in other words, the transmission spectra of N filters have to be ideally orthogonal with zero cross-correlation. For the case of N<<M, the equation (2) is a well-known underdetermined linear algebra problem, which can be solved using linear regression algorithm by minimizing $l_2\ norm$ of the equation (3)(27):

$$minimize\ \|D - SP\|^2\ subject\ to\ 0 \leq P \leq 1 \qquad (4)$$

D, S and P are matrix representation described in equation (3). For the case with stronger measurement noise, regularization of the $l_2$ norm of P or standard deviation of P can be added to the regression with certain weight coefficient $\alpha$ for smooth spectrum reconstruction:

$$\text{minimize } \|D - SP\|^2 + \alpha \|P\|^2 \text{ subject to } 0 \leq P \leq 1 \tag{5}$$

The conceptual schematic of our implementation of a single-shot spectrometer system enabled by broadband filters is shown schematically in Fig. 2(a). In general, such type of spectrometer consists of two stages: splitting of the incident signal into N channels and sampling it by a set of N broadband filters. To split the incident signal into multiple channels, the most straightforward approach is to use a multistage Y-junction tree topology, however, this can result in a large footprint. Besides, the number of output ports is limited to $2^y$, where y refers to the number of stages of Y-junctions for our case y=$\log_2 N$, with a total of N $\log_2 N$ junctions. To reduce the footprint and achieve flexibility in determining the number of channels, we develop a splitter consisting of a series of cascaded N taps attached to a common bus waveguide as shown in Fig.2(b) resulting in the entire footprint of a 32-port splitter as small as 5x256 um$^2$. The design efforts were put to individual tap to ensure small imbalance among the transmission at each channel according to following equation:

$$\kappa_N = \frac{\kappa_0}{1 - (N-1)\kappa_0}, \tag{6}$$

where $\kappa_N$ is the power coupling coefficient at the Nth tap and $\kappa_0$ is the desired transmission coefficient at each port, which for uniform splitting should be $\frac{1}{N}$. The simulated transmission spectra of the coupling regions for a few of the channels are shown in Fig. 2(c). Clearly, each port extracts similar amount of power from the common bus waveguide with very little non-uniformity across the flat regions which can always be accounted for during calibration process.

The core part of the spectrometer is the set of "high-performance" broadband SWFs. The term "high-performance" here is referring to two simultaneously achieved properties: (i) each of SWF should produce a transmission spectrum with diverse features, in other words, the correlation length in the wavelength span (or minimum optical distance between two distinguishable wavelength points) should be small in order to provide high spectral resolution when sampling the input signal spectrum and (ii) the transmission spectra from any two SWFs should be very different (i.e., independent or orthogonal), in order to obtain a sampling matrix $S_{NxM}$ with large rank. It is non-trivial to meet these two requirements while maintaining an ultra-compact footprint, simultaneously. Completely random or disordered medium have been reported to provide diverse spectral features and have been used to demonstrate hyperspectral imaging and ultra-compact spectrometers(28-30). However, for random filters the light will be scattered by numerous defects instead of creating a well-confined, propagating mode in the medium, thereby leading to very high insertion loss for the input signal and consequently result in very low power collected by the detectors. Inspired by this effects, we combine the advantages of disordered medium and high-confinement silicon strip waveguide and develop SWF structure as a broadband filter. Our approach uses controlled perturbations of the Si waveguide by adjusting control parameters (s, w, t) defined in Fig. 2(d). Stratified waveguides or stratified medium has been studied for optical thin film filters, ultrasonic wave propagation, metallic optical fibers, fiber grating couplers, waveguide radiation etc.(31-35). Analogously, the control parameters (s, w, t) determine the effective index and the effective thickness for each layer in the SWF and can be used to design the desired transmission spectrum property of each of the 32 filters. Each layer can be designed to have different parameters in terms of width and length corresponding to their effective indexes and thicknesses. Moreover, the spacing between the consequent layers can also be varied. With our SWF approach we can exploit the large number of these design degrees of freedom to construct efficient broadband spectral transmission filters having diverse spectral features in contrast to previously investigated techniques such as CQD absorption filters and photonics crystal cavities(25, 26). While compared with completely disordered medium, the high confinement from silicon waveguide assures high power collected by the photodetectors.

We use autocorrelation and cross-correlation functions to characterize the filters transmission spectra. The width of the autocorrelation illustrates how "fast and random" are the features in transmission spectrum of the SWF, providing a design metric for achieving the desired resolution. Intuitively, our SWFs transmission functions should have spectral features (i.e., variations) on the same scale as the desired resolution which will be apparent in the width of the main lobe of the corresponding autocorrelation function for each SWF, which therefore can serve as the resolution metric during the design process. The cross-correlation between two distinct SWFs indicates how "different" they are from each other. In ideal case, they should be orthogonal and the cross-correlation should be zero at each point. These desired transmission characteristics of SWFs were achieved by varying the parameters space considering the feature size of current technology, i.e., the parameters s and w are varied in the range of 100nm to 200nm, while h is varying between 100 nm and 300 nm (see definitions in Fig. 2d). The SWF design contains 75-125 layers imposed on a standard silicon strip waveguide (220 nm x 450 nm). All the design parameters s, w, t were varied with an increment step of 5 nm, which is a reasonable assumption for the state of the art CMOS technology. As a consequence, each SWF has an ultra-compact footprint around 1um x 30um, more than 30 times smaller than these reported in previous approaches (26). Simulated transmission spectra of few example SWFs are shown in Fig. 2(e) and Fig. 2(f) for the autocorrelation and the cross-correlation, respectively. Clearly, the spectra of those filters contain diverse features with very little cross-correlation signal. One minor issue with current configuration is the reflections caused by individual SWF. However, the reflected portion of light will only propagate towards the original input port and will not be reflected back again to induce cross-talk among different channels as there doesn't exist any reflective element in the pathway towards input port. In future work, we could implement an identical set of SWFs operating with couplers to the reflected optical signal to recycle these reflected portions of light to further increase the efficiency as well as SNR.

### 3. System validation and characterization

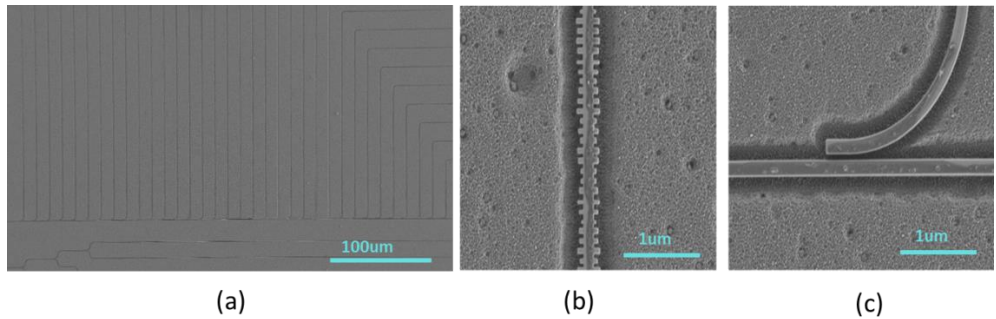

Fig. 3 (a) shows the SEM image of the overview of the entire spectrometer. (b) and (c) gives the zoom-out view of individual SWF as well as the tap based splitter.

The designed spectrometers were fabricated by Applied Nanotools using their standard Multi-Project-Wafer (MPW) tape out. The SEM image of the overview of the entire spectrometer (excluding fiber couplers) is given in Fig. 3(a). And the zoom-out views of the SWF and tap coupler are given in Fig. 3(b-c). Their process uses Electron-beam lithography on a 220nm thick Silicon-On-Insulator (SOI) wafer with a 2-3 um BOX layer, however in our design we intentionally used parameters to also make the fabrication compatible with standard optical lithography. A 2-um oxide protection layer was deposited on top of the silicon structures. Inverted tapers with flat spectral response and less than 5 dB insertion loss were used for fiber to chip coupling. The setup we used for characterization of the chip is a standard optical setup with lensed fibers to couple light in and out of the chip as we are using edge couplers as on-chip fiber couplers to enable broadband operation. The chip is located on a holder with temperature controller and vacuum holder that mechanically stabilizes the chip. The photodetector used in the experiments is an Agilent 81635A power sensor with a sensitivity of -80dBm. As a first step, we measure the transmission spectrum of each filter for calibration, using Agilent 8164B laser with a narrow linewidth output that can be varied from 1460 nm to 1640 nm with a step of about 9 pm (i.e., ~20000 data points in total). The resultant responses for some of the fabricated filters normalized to transmission in a straight waveguide are shown in Fig.4 (a). The dense ripples are due to facet reflections at the fiber couplers. They do not affect the performance significantly due to their relatively small amplitude (less than 0.8dB) compared with the spectra features of the filters' spectra. Also, these measured transmitted signals were used to estimate the auto-correlation for a single filter and cross-correlation of two randomly selected filters (see Fig. 4(b)), which show the expected trends.

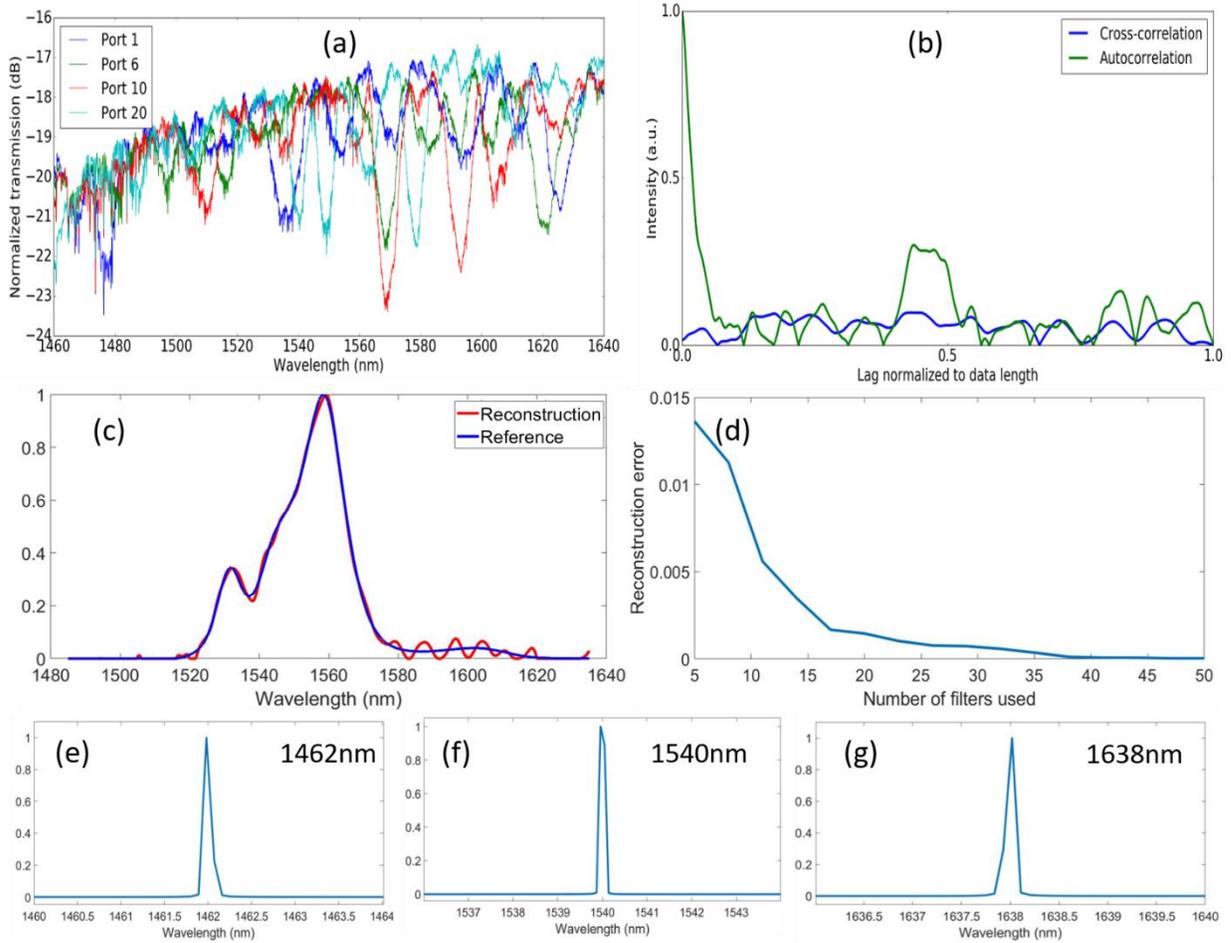

Fig. 4 Experimental validation and characterization of the single-shoot SWF-based spectrometer system: (a) The measured spectra of some SWFs normalized to a straight waveguide transmission; (b) The autocorrelation of a single filter measured spectrum and cross-correlation of two randomly chosen SWFs spectra; (c) Reconstruction of a broadband spectrum using the procedure described by Eqs. 4 and 5; (d) Reconstruction errors as a function of the number of filters used for the input signal spectrum reconstruction; (e-g) Reconstruction of the narrowband spectrum in the spectral range of 1460 nm to 1640 nm. The FWHM of the resolved peak maintains about 0.45nm throughout the 180nm span.

To test the performance of the fabricated SWF spectrometer, we first used a broadband spectrum from ASE source as an input broadband signal with bandwidth of over 60 nm centered around 1550 nm. The output of ASE source was introduced into our spectrometer and the intensity at the output of each filter was detected. To enable the signal processing of the detected signals, each individual SWF spectrum shown in Fig. 4(a) was digitized into a 1-D array containing 450 uniform sampling points in the spectral window from 1485 nm to 1635 nm, such that the incident signal spectrum will be represented by a 1-D array containing 450 unknown coefficients. This number of sampling points can be increased to achieve higher resolution at the cost of longer processing time (Using CVX optimization algorithm implemented in MATLAB environment on a desktop with 4-core processors and 16GB RAM consumes about 0.65s to reconstruct the broadband ASE spectrum). Before performing the reconstruction of the broadband ASE spectrum, the regularization coefficient needs to be pre-calibrated. It is done by sending a broadband superluminescent laser diode (SLD) spectrum with over 80nm spectral components around 1550nm and adjusting the regularization coefficient until a good reconstruction appears. The reconstructed spectrum of the ASE source using the signal processing procedure described in equations 2-5 with calibrated regularization coefficient is shown in Fig 4(c) together with the reference curve obtained by using a commercial Optical Spectrum Analyzer (OSA). The overall reconstruction is found in good agreement with the reference spectrum except the low-amplitude tail in the optical range longer than 1580 nm. This occurs due to the fact that the spectral components with higher amplitude have higher weights than the components with lower amplitude when solving equation (2). This is a common disadvantage for any multiplexing spectrometers. However, this deficiency can be effectively overcome by using larger number of filters or a priori knowledge about the constraints on the incident signal or using segmented spectrometers, each of which only reconstructs a sub-band of the incident signal. In order to determine how many filters are necessary for broadband spectrum reconstruction, we also fabricated 50 stand-alone SWFs. The reconstruction errors as a function of number of filters used for spectrum reconstruction is shown in Fig. 4(d). Clearly, using more filters leads to better quality of reconstruction, but at the expense of lower input power (i. e., larger number of channels) and, consequently, lower SNR.

Next, we test the capability to reconstruct narrowband components of the SWF spectrometer by sending narrow linewidth spectra from a tunable laser source. For the narrowband spectrum reconstruction, the corresponding regularization coefficient is calibrated by sending peaks at 10 different locations (from 1460nm till 1640nm) and adjust the corresponding coefficient for best reconstruction results. Then the average of the 10 values is set as the common regularization coefficient for narrowband spectral components. In total, we performed over 10 measurements with varying wavelengths. Three examples with spectral lines at 1462 nm, 1540 nm and 1638 nm are given in Fig. 4(e-g). The achieved FWHM of the resolved peak at different wavelength locations maintains about 0.45nm within 180nm optical range. The reconstruction results shown in Fig.4 (e-g) confirm the high spectral resolution as well as the broad operation bandwidth from 1460 nm to 1640 nm, which is purely limited by our measurement capability.

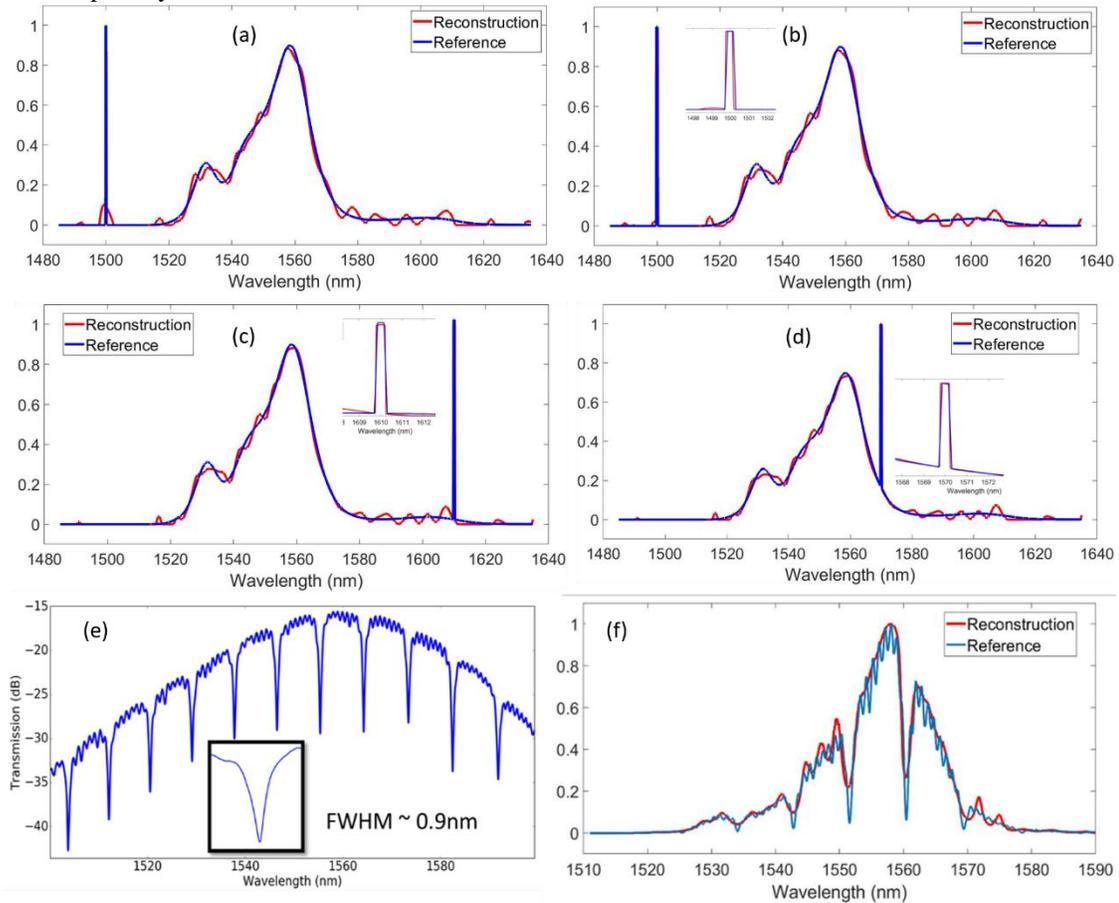

Fig. 5 Experimental validation and characterization of the single-shoot SWF-based spectrometer for hybrid spectra that contain both broad spectral components (e. g., ASE source) and narrow spectral peak or notches: (a) Results obtained without further optimization of the reconstruction algorithm using Eqs. 4-5 showing poor quality in revealing the narrow spectral peak signals; (b-d) Results obtained from using advanced reconstruction algorithm of Eq. 7 showing good reconstruction quality of all the three narrow spectral peak signals at different locations. Insets in (b-d) show the zoom in area around the narrow line signals.(e) shows the notches generated by a ring resonator at a different chip. (f) plots the reconstruction of the broadband spectrum with narrow notches. Even if the performance compared degrades compared with reconstruction of pure broadband spectrum, but the individual notches and the overall envelope can be accurately captured.

Besides, we also investigate how the spectrometer can handle mixed spectrum, where both broad band spectrum and narrow band spectrum are present. To synthesize such an input signal, we use a 3dB coupler to combine the output from the narrow linewidth tunable laser source and the output from the C+L band ASE (broadband spectrum) sources. This combined signal is then introduced to the input of the spectrometer. As mentioned above, the regularization weight $\alpha$ in equation (5) is different for reconstruction of a broadband spectrum signal and narrowband spectrum signal. Therefore, if we use the term $\alpha$ optimized for broadband spectrum, the reconstruction quality of the narrow band signal peak is poor as shown in Fig. 4(a). However, we can use segmented regularization in equation (5) with different weights $\alpha$ for broad and narrow spectral components:

$$minimize\ \|D - SP\|^2 + \alpha_1 \|P_1\|^2 + \alpha_2 \|P_2\|^2\ subject\ to\ 0 \leq P_{1,2} \leq 1, \quad (7)$$

where the to-be-reconstructed 1-D array P is separated into two parts $P_1$ and $P$, which represent the narrow and broad spectral components respectively, and $\alpha_1, \alpha_2$ refer to the optimized regularization weights for narrow and broad spectral components reconstruction, respectively. The two terms $\alpha_1\ and\ \alpha_2$ can be accurately pre-determined using extra calibration procedure by sending known narrowband and broadband signal into the spectrometer separately and adjust the corresponding weight term. Consequently, with this approach, the overall input spectrum signal can be reconstructed with good quality as demonstrated by the reconstructed results of spectra consisting of ASE spectrum and narrow spectrum peak at three different locations (see Fig. 4(b-d)).

Finally, we take the challenge of sending broadband spectrum with narrow notches to further test the capability of our spectrometer. This kind of signal is generated by first sending the ASE spectrum to a separate chip that has ring resonators to introduce notches. This chip is located in a separate testing setup that uses grating couplers and standard single-mode fibers as fiber/chip interface, then the output from the chip with the ring is sent to the spectrometer chip which is located in a different testing setup that uses edge couplers and lensed fibers to allow broad band testing. The measured transmission spectrum of the ring resonator using tunable laser source is plotted in Fig, 5(e), showing that the fabricated ring resonator has a measured FWHM of about 0.9nm and FSR of about 9 nm. The ripples in Fig. 5(e) are due to parasitic reflections at the grating couplers as well as fiber facets. The measured transmission of the ASE source through the ring resonator chip using a commercial OSA is shown in Fig. 5(f) (blue curve) and the corresponding reconstruction obtained from the spectrometer chip is shown in Fig. 5(f) (red curve). Clearly, the individual notch can be accurately reconstructed. Compared with reconstructions of narrow peaks and broadband spectrum, the mismatch in this case is clearer, as for this type of an underdetermined problem, solving a sparse matrix (i. e., corresponding to a spectrum only containing narrow peaks ) is well-known to be easier than dense matrix (i.e., corresponding to a broadband spectrum with narrow notches).

Note that, the measurement and characterization errors also impact the spectrometer performance. The potential contributions to the errors could be temperature variation between device calibration and spectrum reconstruction, as it will cause shift of the filters' transmission spectrum and lead to inaccurate preparation of the sampling matrix (S in equation (3)) when performing spectrum reconstructions. We use a temperature controller to avoid this issue. Mechanical instability of the chip and the fibers could also contribute to the measurement errors. We use a vacuum pump to stabilize the chip, and fiber holder to stabilize the fibers. The setup is also pre-calibrated to ensure no fiber drift takes place within reasonable measurement time. In real applications, the spectrometer could be packaged with fiber arrays as optical I/O to fully avoid these instabilities.

## 4. Conclusion

In this manuscript, we implement a concept of single-shot spectrometer that uses a novel broadband SWFs realized in silicon photonics platform. The 32 SWFs are utilized as the broadband filters to sample the incident signal exploiting the multiple design degrees of freedom that enable design and fabrication of spectral filters with diverse spectral features that possess narrow autocorrelation for high resolution and low cross-correlation between the different filters in the array to increase orthogonality for effective signal reconstruction. In order to distribute the incident spectral signal into 32 filters with little power imbalance, we also developed and demonstrated experimentally an ultra-compact splitter based on cascaded taps. The total footprint of the splitter and the 32 SWFs is as small as 35um x 260um, which to the best of our knowledge, is the smallest experimentally demonstrated spectrometer on silicon photonic platform. The experimental results demonstrate operation with broad bandwidth input signals (i.e., 180 nm centered at 1550 nm), narrow band signals (i.e., 0.45 nm FWHM laser emission) and mixed broad/narrow band input signals. The resolution can be further increased by optimizing the SWFs with even sharper auto-correlation and smaller cross-correlation or by combining other spectrum retrieval techniques such as compressive sensing approach. While the concept is demonstrated in the optical range around 1550 nm, its realization can be easily extended to other material platform operating in the other desired optical spectral (e.g., SiN for operation in visible and mid infrared range). The SWF spectrometer approach is a promising candidate for cost effective manufacturing of miniaturized spectrometers making them suitable candidate for integration with various mobile and portable systems.

*Funding*

This work was supported by the Defense Advanced Research Projects Agency (DARPA) DSO NLM and NAC Programs, the Office of Naval Research (ONR), the National Science Foundation (NSF) grants DMR-1707641, CBET-1704085, NSF ECCS-180789, NSF ECCS-190184, NSF ECCS-2023730, the Army Research Office (ARO), the San Diego Nanotechnology Infrastructure (SDNI) supported by the NSF National Nanotechnology Coordinated Infrastructure (grant ECCS-2025752), the Quantum Materials for Energy Efficient Neuromorphic Computing-an Energy Frontier Research Center funded by the U.S. Department of Energy (DOE) Office of Science, Basic Energy Sciences under award # DE-SC0019273.Advanced Research Projects Agency-Energy (LEED: A Lightwave Energy-Efficient Datacenter), and the Cymer Corporation.